15/VII/2013

**X-ray diffraction by magnetic charges (monopoles)**

S W Lovesey[1,2] and D D Khalyavin[1]

1. ISIS Facility, STFC Oxfordshire OX11 0QX, UK

2. Diamond Light Source Ltd, Oxfordshire OX11 0DE, UK

**Abstract** Magnetic charges, or magnetic monopoles, may form in the electronic structure of magnetic materials where ions are deprived of symmetry with respect to spatial inversion. Predicted in 2009, the strange magnetic, pseudo-scalars have recently been found different from zero in simulations of electronic structures of some magnetically ordered, orthorhombic, lithium orthophosphates ($LiMPO_4$). We prove that magnetic charges in lithium orthophosphates diffract x-rays tuned in energy to an atomic resonance, and to guide future experiments we calculate appropriate unit-cell structure factors for monoclinic $LiCoPO_4$ and orthorhombic $LiNiPO_4$.

## 1. Introduction

Absence of spatial inversion symmetry on a local scale in the electronic structure of a magnetic material allows the formation of atomic magnetic charge [1]. This unusual entity is a scalar (monopole) magneto-electric multipole that is parity-odd and time-odd [2]. Magnetic charge contributes to resonant x-ray Bragg diffraction but not the corresponding dichroic signals [1, 3]. There is solid experimental evidence for magneto-electric dipoles (anapoles) in vanadium sesquioxide [4, 5] and haematite [6]. Evidence for magnetic charges in gallium ferrate is less convincing [7], and the prediction for its existence in haematite is not tested [6]. Interest in magneto-electric multipoles, and counterpart polar multipoles that are parity-odd and time-even, stems from their practical importance, in device design and demanding tests of simulations of the electronic structure of complex materials, and novelty, of course.

Observation of magneto-electric multipoles requires a test with matching discrete symmetries, and x-ray absorption and scattering are candidate tools. Parity-odd absorption derived from electric dipole - magnetic dipole (E1-M1) and electric dipole - electric quadrupole (E1-E2) events give access to magneto-electric and polar multipoles with rank 0 - 2 and 1 - 3, respectively [1, 8]. Thus, anapoles contribute in both E1-M1 and E1-E2 events, whereas magnetic charge is found in E1-M1 only [9, 10].

Lithium orthophosphates of divalent manganese (lithiophilite), iron (triphylite), cobalt, and nickel are members of the olivine family of structures (some minerals in the family of structures possess the colour olive-green), with magnetic motifs indexed on the chemical unit cell (Γ-point) [11]. Cell dimensions of lithium orthophosphates permit Bragg diffraction at L-edges, which are 2p → 3d dipole-allowed in the soft x-ray region with a wavelength $\lambda \approx 14$ Å. X-ray diffraction is strongly enhanced by $L_{2,3}$ edges of 3d transition metal ions and may display a weak E1-M1 event and, in consequence, magnetic charge [7, 12, 13].

In this communication, we report electronic structure factors for transition metal ions in lithium orthophosphates that describe diffraction, of both neutrons and x-rays, and dichroic signals. Results make use of established chemical and magnetic symmetries. Our findings with regard to magnetic charge accord with *ab initio* simulations of electronic structure in orthorhombic crystals [14]. It is shown here that, x-ray diffraction by $LiNiPO_4$ can confirm magnetic charge predicted in simulations of this compound. A second example of this kind is $LiCoPO_4$. For this compound magnetic charge is allowed in the monoclinic structure proposed by Vaknin et al [15], while it is forbidden by symmetry in the undistorted, orthorhombic structure used in [14].

Orthorhombic, magnetic lithium orthophosphates are discussed in Sections 2, 3 and 4. In the next section their point symmetries and multipoles, followed, in Section 3, by electronic structure factors for bulk properties, neutron diffraction and x-ray diffraction. Resonant x-ray Bragg diffraction is the subject of Section 4. The same topics are covered in Section 5 for the monoclinic compound $LiCoPO_4$. Conclusions are gathered in Section 6.

## 2. Orthorhombic space groups and point-symmetry

The chemical structure is orthorhombic Pnma-type with metal ions using sites (4c) that are not centres of inversion symmetry. Crystal class $D_{2h}$ (mmm) is centrosymmetric.

There are three magnetic motifs of interest for lithium compounds $LiMPO_4$ with M = Mn, Fe, Co and Ni, and we label them by the corresponding parity-odd, one-dimensional irreducible representations (irrep) that are $\Gamma_1$, $\Gamma_2$, and $\Gamma^-_4$. Here we list the motifs, space-groups, crystal class, and point symmetry of sites (4c) used by metal ions. (Magnetic space-groups are specified in terms of the Miller and Love and Belov-Neronova-Smirnova notations. [16, 17])

$\Gamma^-_1$ : ($m_a$, 0, $m_c$) (–$m_a$, 0, $m_c$) (–$m_a$, 0, –$m_c$) ($m_a$, 0, $m_c$) Mn ions,

Pn'm'a', m'm'm', $m_y$'

$\Gamma^-_2$ : (0, $m_b$, 0) (0, –$m_b$, 0) (0, –$m_b$, 0) (0, $m_b$, 0) Fe & Co ions,

Pnma', mmm', $m_y$

$\Gamma^-_4$ : ($m_a$, 0, $m_c$) ($m_a$, 0, –$m_c$) (–$m_a$, 0, –$m_c$) (–$m_a$, 0, $m_c$) Ni ions,

Pnm'a, mm'm, $m_y$'. (2.1)

Magnetic crystal classes that appear here occur in the 58 classes that allow a linear magneto-electric effect.

Electronic degrees of freedom of the magnetic ions are encapsulated in irreducible, spherical multipoles. Our notation for a generic spherical multipole is $\langle O^K_Q \rangle$, with a complex conjugate $\langle O^K_Q \rangle^* = (-1)^Q \langle O^K_{-Q} \rangle$, where the positive integer K is the rank and Q the projection, which satisfies $-K \leq Q \leq K$ [8, 18]. Angular brackets $\langle ... \rangle$ denote the time-average of the enclosed quantum-mechanical operator, i.e., a multipole is a property of the electronic ground-state. Parity-even multipoles are denoted by $\langle T^K_Q \rangle$ and they have a time signature $\sigma_\theta = (-1)^K$. In x-ray scattering, they arise in a description of Thomson scattering and resonant Bragg diffraction enhanced by a parity-even event, e.g., E1-E1 and E2-E2 events. They arise also in a description of neutron diffraction [8]. In this case, K is odd, and for nd-ions K has has a maximum value 5 (triakontadipole) and for nf-ions K has a maximum value 7 (octaeicosahecatontapole). For parity-odd events in x-ray diffraction our notation for multipoles is $\langle U^K_Q \rangle$ for time-even, polar and $\langle G^K_Q \rangle$ for time-odd, magneto-electric [8, 18]. The time-odd pseudo-scalar $\langle G^0_0 \rangle$ is magnetic charge, $\langle G^1 \rangle$ is an anapole (toroidal dipole), and the time-even pseudo-scalar $\langle U^0_0 \rangle$ is chirality (helicity). The five standard dichroic signals (linear, circular, natural circular, magneto-chiral and non-reciprocal linear) can be written in terms of multipoles, and expressions for them are found in reference [3].

Magnetic charge can be represented by the scalar product of spin, **S**, and position, **R**, operators with $\langle G^0_0 \rangle = \langle \mathbf{S} \cdot \mathbf{R} \rangle$ [3, 9]. It is allowed in an E1-M1 event, where it contributes to diffraction in which polarization is rotated, namely, σ→π and π→σ (Figure 1) [1]. The physical content of multipoles in resonant Bragg diffraction is brought out in the sum-rules they obey. For diffraction enhanced by L-edges, parity-even dipoles (K = 1) satisfy,

$$\langle \mathbf{T}^1 \rangle_{L2} + \langle \mathbf{T}^1 \rangle_{L3} = - \langle \mathbf{L} \rangle_{3d}/(10\sqrt{2}),$$

where $\langle \mathbf{L} \rangle_{3d}$ is orbital angular momentum in the 3d-valence state, and quadrupoles (K = 2) satisfy,

$$\langle \mathbf{T}^2 \rangle_{L2} + \langle \mathbf{T}^2 \rangle_{L3} = \langle \{\mathbf{L} \otimes \mathbf{L}\}^2 \rangle_d/60,$$

where the tensor product $\{\mathbf{L} \otimes \mathbf{L}\}^2$ has for its diagonal component (Q = 0) a value $\{\mathbf{L} \otimes \mathbf{L}\}^2_0 = [3L^2_z - L(L + 1)]/\sqrt{6}$, which demonstrates affinity to a standard, parity-even quadrupole operator. For magneto-electric dipoles one finds,

$$\langle \mathbf{G}^1 \rangle_{L2} + \langle \mathbf{G}^1 \rangle_{L3} = - \langle [\mathbf{R} \times (\mathbf{L} + 2\mathbf{S})] \rangle_{3d}/(2\sqrt{2}),$$

which demonstrates that $\langle \mathbf{G}^1 \rangle$ is related directly to standard spin and orbital anapoles.

2.1 consequences of point-group symmetry

There are two site symmetries to consider, namely, $m_y$ and $m_y'$. Site symmetry is implemented using $2_y \langle O^K_Q \rangle = (-1)^K \langle O^K_Q \rangle^*$.

The identity $m_y \langle O^K_Q \rangle = \langle O^K_Q \rangle$ leads to the condition,

$$\sigma_\pi (-1)^{K+Q} \langle O^K_{-Q} \rangle = \sigma_\pi (-1)^K \langle O^K_Q \rangle^* = \langle O^K_Q \rangle, \qquad (2.2)$$

where $\sigma_\pi = \pm 1$ is the parity-signature of $\langle O^K_Q \rangle$. The magnetic dipole, $\sigma_\pi = + 1$, $\langle \mathbf{T}^1 \rangle$ points along the b-axis, as expected. Parity-odd ($\sigma_\pi = -1$) multipoles with K odd are purely real. In particular, the c-axis component of the anapole, $\langle G^1_0 \rangle$, is allowed and magnetic charge is not allowed, $\langle G^0_0 \rangle = 0$, for point-group symmetry $m_y$.

The identity $m_y'\langle O^K_Q \rangle = \langle O^K_Q \rangle$ leads to quite different physical properties. In place of (2.2) we have,

$$\sigma_\theta \sigma_\pi (-1)^{K+Q} \langle O^K_{-Q} \rangle = \sigma_\theta \sigma_\pi (-1)^K \langle O^K_Q \rangle^* = \langle O^K_Q \rangle, \qquad (2.3)$$

where $\sigma_\theta = \pm 1$ is the time-signature. Parity-even ($\sigma_\pi = +1$) multipoles have $\sigma_\theta = (-1)^K$, and $\langle T^K_Q \rangle = \langle T^K_Q \rangle^*$ from (2.3), i.e., all parity-even multipoles are purely real and the magnetic dipole is confined to the a-c plane. The result $\sigma_\theta \sigma_\pi = (-1)(-1) = +1$ applies for magneto-electric multipoles, and the requirement $\langle G^K_Q \rangle = (-1)^K \langle G^K_Q \rangle^*$, means magnetic charge is allowed while $\langle G^K_0 \rangle = 0$ for K odd, in point-group symmetry $m_y'$.

## 3. Orthorhombic structure factors

Unit-cell structure factors for scattering are derived from [8],

$$\Psi^K_Q = \sum_d \exp(i\mathbf{d}\cdot\mathbf{\tau}) \langle O^K_Q \rangle_d, \qquad (3.1)$$

where the sum is over the four magnetic ions at positions $\mathbf{d}$ in the unit cell, and the Bragg wavevector $\mathbf{\tau} = (h, k, l)$ with integer Miller indices. Using sites (4c) at: x, 1/4, z; – x +1/2, 3/4, z + 1/2; – x, 3/4, – z; x + 1/2, 1/4, – z + 1/2;

$$\Psi^K_Q = \exp(i\varphi)\langle O^K_Q \rangle_1 + (-1)^{h+l}\exp(-i\varphi')\langle O^K_Q \rangle_2$$

$$+ \exp(-i\varphi)\langle O^K_Q \rangle_3 + (-1)^{h+l}\exp(i\varphi')\langle O^K_Q \rangle_4, \qquad (3.2)$$

with spatial phase-factors $\varphi = 2\pi(hx + k/4 + lz)$ and $\varphi' = 2\pi(hx + k/4 - lz)$ [19]. Environments at sites 2, 3, and 4 are related to site 1 by rotations by 180° about crystal axes, and they are $2_z$, $2_y$, and $2_x$, respectively. In ensuing calculations, we will use $2_z \langle O^K_Q \rangle = (-1)^Q \langle O^K_Q \rangle$, and $2_x \langle O^K_Q \rangle = (-1)^Q 2_y \langle O^K_Q \rangle$.

### 3.1 charge diffraction

Pure charge diffraction by the magnetic ions is created by parity-even and time-even multipoles, with $\sigma_\theta = (-1)^K = +1$ for K even. Corresponding multipoles are purely real for both sites symmetries and,

$$\Psi^K_Q(\text{charge}) = 2\langle T^K_Q \rangle [\cos\varphi + (-1)^{h+l+Q}\cos\varphi'], \qquad (3.3)$$

which is purely real, for the chemical structure is centrosymmetric. Standard extinction rules for Miller indices satisfy $\Psi^K_0(\text{charge}) \neq 0$, whereas Templeton & Templeton scattering, created by angular anisotropy in electronic structure, is labelled by Miller indices and $Q \neq 0$. Thus, to avoid charge scattering from orthorhombic structures consider reflections (h, k, 0) with h odd that give $\Psi^K_0(\text{charge}) = 0$.

3.2 magnetic diffraction

All three structure factors are proportional to $\langle O^K_Q \rangle$ which is subject to either (2.2) or (2.3). Structure factors for $\Gamma^-_1$ and $\Gamma^-_2$ are distinguished solely by point-group symmetry, i.e., properties of $\langle O^K_Q \rangle$, and for this reason we only write out $\Psi^K_Q(\Gamma^-_1)$. From (3.2),

$$\Psi^K_Q(\Gamma^-_1) = \langle O^K_Q \rangle [\exp(i\varphi) + \sigma_\theta \sigma_\pi \exp(-i\varphi)$$

$$+ (-1)^{h+l+Q} (\sigma_\theta \sigma_\pi \exp(i\varphi') + \exp(-i\varphi'))]$$

$$= \langle O^K_Q \rangle [\exp(i\varphi) + \sigma_\theta \sigma_\pi \exp(-i\varphi)] [1 - \sigma_\theta \sigma_\pi (-1)^Q]; \ (h, k, 0) \text{ with } h \text{ odd}, \quad (3.4)$$

$$\Psi^K_Q(\Gamma^-_4) = \langle O^K_Q \rangle [\exp(i\varphi) + \sigma_\theta \sigma_\pi \exp(-i\varphi)$$

$$+ \sigma_\theta (-1)^{h+l+Q} (\sigma_\theta \sigma_\pi \exp(i\varphi') + \exp(-i\varphi'))]$$

$$= \langle O^K_Q \rangle [\exp(i\varphi) + \sigma_\theta \sigma_\pi \exp(-i\varphi)] [1 - \sigma_\pi (-1)^Q]; \ (h, k, 0) \text{ with } h \text{ odd}. \quad (3.5)$$

Time-odd multipoles vanish in the paramagnetic phase. Second equalities in (3.4) and (3.5) apply for x-ray reflections at which charge scattering is forbidden, and describe Templeton and Templeton scattering in the paramagnetic phase.

The expressions (3.4) and (3.5) apply to bulk properties, x-ray diffraction, and neutron diffraction. By way of an example of neutron diffraction, we consider $LiMnPO_4$, illustrated in Figure 2, and find $\Psi^1_0(\Gamma^-_1) = 0$ (c-axis) and $\Psi^1_1(\Gamma^-_1) \neq 0$ (a-axis) for reflections (0, 1, 0), (0, 1, 2) and (2, 3, 0), results which agree with observations at 2 K [20]. Magnetic neutron diffraction is bound by a selection rule; if all electronic dipoles are parallel with the Bragg wavevector $\tau = (h, k, l)$ the intensity is zero, no matter what value the magnetic structure factor takes.

3.3 bulk properties

Bulk properties of magnetic ions are determined by structure factors evaluated for Miller indices $h = k = l = 0$. For this condition,

$$\Psi^K_Q(\Gamma^-_1) = \Psi^K_Q(\Gamma^-_2) = \langle O^K_Q \rangle_\Gamma [1 + \sigma_\theta \sigma_\pi] [1 + (-1)^Q], \text{ Bulk} \quad (3.6)$$

and the expressions applies to orthorhombic crystals containing Mn, Fe or Co ions. Magnetic dipoles, and all other parity-even and time-odd multipoles, possess $\sigma_\theta \sigma_\pi = -1$ and make no contribution to bulk properties. The same is true of magnetic charge for $\Gamma^-_2$, because $\langle G^0_0 \rangle = 0$ by virtue of the point-group symmetry, $m_y$. But there is a net magnetic charge for $\Gamma^-_1$, together with all other magneto-electric multipoles with Q even and not forbidden by $\langle G^K_0 \rangle = 0$ for K odd in point-group symmetry $m_y'$. Turning to $\Gamma^-_4$,

$$\Psi^K_Q(\Gamma^-_4) = \langle O^K_Q \rangle [1 + \sigma_\theta \sigma_\pi] [1 + \sigma_\theta (-1)^Q], \text{ Bulk} \qquad (3.7)$$

for Ni ions. Here, also, magnetic moments fully compensate, together with all other parity-even magnetic multipoles. There is no net magnetic charge, and a net anapole moment resides along the b-axis [21, 22].

## 4. Orthorhombic unit-cell structure factors

Unit-cell structure factors for x-ray diffraction, F, are very easily derived using $A_{K,Q} = A_{K,-Q} = (\Psi^K_Q + \Psi^K_{-Q})/2$ with $A_{K,0} = \Psi^K_Q$, and $B_{K,Q} = - B_{K,-Q} = (\Psi^K_Q - \Psi^K_{-Q})/2$ and $B_{K,0} = 0$. Unit-cell structure factors for E1-E1 and E1-M1 events are found in references [1, 23]. $A_{K,Q}$ and $B_{K,Q}$, are calculated relative to states of polarization and a Bragg wavevector (h, k, l) shown in Figure 1; specifically, the setting of the crystal at the origin of a rotation of the crystal by an angle $\psi$ around the Bragg wavevector (azimuthal-angle scan).

$\Gamma^-_1$: From (3.4) we see that diffraction at reflections (h, k, 0) with h odd from magneto-electric multipoles, $\sigma_\theta \sigma_\pi = + 1$, is allowed for Q odd. In consequence, magnetic charge, $\langle G^0_0 \rangle$, is not observed.

$\Gamma^-_2$: Site symmetry $m_y$ does not allow magnetic charge. In Section 5 we examine consequences of a lower symmetry structure for LiCoPO$_4$ proposed by Vaknin et al [15] and find that magnetic charge is allowed and contributes to diffraction at space-group forbidden reflections.

$\Gamma^-_4$: Unit-cell structure factors for (h, k, 0) with h odd are derived from the second equality in (3.5). They are purely real for each of the four polarization channels, which mean (a) diffraction is independent of circular polarization in the primary beam, and (b) unit-cell structure factors add coherently, unless mixing parameters are complex and depend on energy.

The Bragg angle is determined by $\sin\theta = (\lambda/2)|(h, k, l)|$ and $\lambda \approx 14.4$ Å at the Ni L$_3$ absorption edge (a = 10.02 Å, b = 5.83 Å, c = 4.66 Å [20]). Whence, diffraction is allowed at (1, 0, 0). Chemical and magnetic structures of LiNiPO$_4$ are depicted in Figure 2.

The Bragg wavevector (h, k, 0) is aligned with − x in Figure 1 by rotation by an angle α about the c-axis (z-axis). We find,

$$A_{K,Q} = (1/2) \Psi^K_Q(\Gamma^-_4)[\exp(-iQ\alpha) + \sigma_\theta \sigma_\pi (-1)^{K+Q} \exp(iQ\alpha)]$$

$$B_{K,Q} = (1/2) \Psi^K_Q(\Gamma^-_4)[\exp(-iQ\alpha) - \sigma_\theta \sigma_\pi (-1)^{K+Q} \exp(iQ\alpha)]. \quad (4.1)$$

For the particular case of (h, 0, 0) α = π. For this reflection, there is no diffraction in channels with unrotated polarization, $F_{\pi'\pi} = F_{\sigma'\sigma} = 0$, while for the channel with rotated polarization $F_{\pi'\sigma} = F_{\sigma'\pi}$ and,

$$F_{\pi'\sigma} = 4\cos(\varphi)\,[\sin(\theta)\tan(\varphi)\langle T^1_{+1}\rangle + \cos(\theta)\cos(\psi)\langle T^2_{+1}\rangle$$

$$+ (1/\sqrt{2})\sin(2\theta)\cos(\psi)\tan(\varphi)\langle U^1_0\rangle - (2/\sqrt{3})\sin^2(\theta)\langle G^0_0\rangle$$

$$- (1/2\sqrt{6})[2 + \cos^2(\theta)(1 + 3\cos(2\psi))]\langle G^2_0\rangle$$

$$+ (1/2)[2 + \cos^2(\theta)(1 - \cos(2\psi))]\langle G^2_{+2}\rangle]. \quad (4.2)$$

The contribution to diffraction from magnetic charge does not change with rotation of the crystal about the Bragg wavevector by an angle ψ, as might be anticipated. The time-odd multipoles, $\langle T^1_{+1}\rangle$ and $\langle G^K_Q\rangle$, vanish in the absence of long-range magnetic order, and Templeton and Templeton scattering is created by the quadrupole $\langle T^2_{+1}\rangle$.

Chemical and magnetic symmetry, together with discrete symmetries of multipoles, are the foundation of expression (4.2) for resonant x-ray Bragg diffraction. Resonant energy denominators are conspicuous by their absence. This is a singular benefit of the fast-collision approximation, which amounts to neglect of angular anisotropy in the wavefunction for the intermediate state that accepts the photon [8]. Loss of the benefit results in a vastly more complicated expression for the scattering length, the starting point for an electronic structure factor, that might not be justified [24, 25].

Contributions in (4.2) depend on the x-ray wavelength and radial integrals, which are not made explicit. To create a meaningful measure of relative strengths of E1-E1 and E1-M1 contributions it is useful to consider a dimensionless quantity ρ that includes q = 2π/λ and atomic radial integrals. In the case of an E1-M1 event, there is the familiar dipole radial integral from the E1-event, namely, (Θ| R |Ξ) where Θ is a valence state, and Ξ is the intermediate state which accepts the photon. The second radial integral, (Θ'| Ξ), is the radial

part of the matrix element of the magnetic moment, and the valence state Θ' and Ξ have identical orbital angular momentum, l', because the magnetic moment operator does not change angular momentum. With these definitions, a value of ρ(E1-M1) is,

ρ(E1-M1) = q (Θ| R| Ξ) (Θ'| Ξ).

Likewise, for an E1-E1 event,

ρ(E1-E1) = [(Θ| R| Ξ)/$a_o$]$^2$(mΔ$a_o^2$/$\hbar^2$),

which has no explicit dependence on q. In the dimensionless factor (mΔ$a_o^2$/$\hbar^2$) the Bohr radius is $a_o$ and Δ is the energy of the resonance.

## 5. LiCoPO$_4$

Vaknin et al [15] show that in LiCoPO$_4$ the Co magnetic dipole departs from the b-axis and the appropriate magnetic space-group is monoclinic P2$_1$'11 (unique axis a), crystal class 2$_x$', and not Pnma1'. Here, we demonstrate that magnetic charge is allowed and contributes to diffraction for this monoclinic structure. The chemical and magnetic structures (Γ$_2$) of LiCoPO$_4$ are depicted in Figure 3.

Initial positions (4c) in Pnma split using two non-equivalent groups;

(2a: 1) (x, 1/4, z; x + 1/2, 1/4, − z + 1/2),

and,

(2a: 2) (− x + 1/2, 3/4, z + 1/2; − x, 3/4, − z).     (5.1)

The site symmetry 1 (identity) allows magnetic charge, of course. Two environments inside (2a) are related by the 2$_x$' symmetry element. We go on to find,

Ψ$^K_Q$(1) = exp(iφ)[⟨O$^K_Q$⟩ + σ$_θ$ (− 1)$^{h + l + K}$ exp(− 4πilz) ⟨O$^K_{-Q}$⟩],     (5.2)

Ψ$^K_Q$(2) = exp(− iφ)[(− 1)$^{h + l}$ exp(4πilz)⟨O$^K_Q$⟩ + σ$_θ$ (− 1)$^K$ ⟨O$^K_{-Q}$⟩],     (5.3),

where φ = 2π(hx + k/4 + lz), as before. Note the absence in (5.2) and (5.3) of an explicit dependence on the parity of multipoles. As in previous cases, the electronic structure factors are appropriate for neutron and x-ray scattering.

In the nominal setting of the crystal orthogonal axes (**a**, **b**, **c***) match (x, y, z) in Figure 1. Here, **a** = a (1, 0, 0), **b** = b (0, 1, 0), **c** = c (0, cos$\alpha_o$, sin$\alpha_o$) with cell volume $v_o$ = abc sin$\alpha_o$. For a reflection (h, k, 0) the Bragg wavevector $\tau$ = 2$\pi$(h/a, k/b, – (k/b) cot$\alpha_o$).

Bulk properties are determined by,

$$\Psi^K_Q(1) = \Psi^K_Q(2) = [\langle O^K_Q \rangle + \sigma_\theta (-1)^K \langle O^K_{-Q} \rangle]. \text{ Bulk}$$

Evaluation of this quantity for the magnetic dipole, $\langle \mathbf{T}^1 \rangle$, and the anapole, $\langle \mathbf{G}^1 \rangle$, show that both have net values in the b-c* plane. Unlike orthorhombic structures, net magnetization can be different from zero. A net polar dipole resides along the a-axis. There is no net magnetic charge. A linear magneto-electric effect is allowed, and with 2'11 (unique axis a) setting $\alpha_{12}$, $\alpha_{13}$, $\alpha_{21}$, $\alpha_{31}$ are components of the tensor allowed different from zero.

X-ray charge scattering is absent for h odd and l = 0, the same condition as in Pnma, and electronic structure factors reduce to,

$$\Psi^K_Q(1) = \exp(i\varphi)[\langle O^K_Q \rangle - \sigma_\theta (-1)^K \langle O^K_{-Q} \rangle], \quad (5.4)$$

for space-group reflections (h, k, 0) with h odd, with a similar expression for $\Psi^K_Q(2)$. Applying (5.4) to parity-even (t), polar (u) and magneto-electric (g) multipoles,

$$\Psi^K_Q(t) = -\Psi^K_{-Q}(t) \text{ and } \Psi^K_0(t) = 0, \quad (5.5)$$

$$\Psi^K_Q(u) = -(-1)^K \Psi^K_{-Q}(u) \text{ with } \Psi^K_Q(u) = 0 \text{ for K even,}$$

and $\Psi^1_{+1}(u) = -i\sqrt{2} \exp(i\varphi)\langle U^1_b \rangle, \quad (5.6)$

$$\Psi^K_Q(g) = (-1)^K \Psi^K_{-Q}(g) \text{ with } \Psi^1_0(g) = 0,$$

and $\Psi^1_{+1}(g) = -\sqrt{2} \exp(i\varphi)\langle G^1_a \rangle. \quad (5.7)$

It follows from (5.5) that the magnetic dipole parallel to the c*-axis does not contribute to diffraction and charge scattering is absent, as expected. Contributions from polar dipoles are parallel to the b-axis and c*-axis, and the anapole contribution is parallel to the a-axis.

We calculate $A_{K,Q}$ and $B_{K,Q}$ with an analogue of (4.1) and find,

$$A_{K,Q} = (1/2)[\exp(-iQ\pi) \Psi^K_Q + \exp(iQ\pi) \Psi^K_{-Q}],$$

$$B_{K,Q} = (1/2)[\exp(-iQ\pi)\Psi^K_Q - \exp(iQ\pi)\Psi^K_{-Q}], \quad (5.8)$$

for the class of reflection (h, 0, 0) with h odd. Expression (5.8) is used in conjunction with (5.5), (5.6) and (5.7). Magnetic charge contributes in the rotated channel of diffraction and we give the appropriate structure factor,

$$F_{\pi'\sigma} = F_{\sigma'\pi} = i\sin(\theta)B_{1,1}(t) - \cos(\theta)\cos(\psi)B_{2,1}(t) + i\cos(\theta)\sin(\psi)B_{2,2}(t)$$

$$- (i/\sqrt{2})\sin(2\theta)\cos(\psi)A_{1,0}(u) - \sin(2\theta)\sin(\psi)A_{1,1}(u)$$

$$- (2/\sqrt{3})\sin^2(\theta)A_{0,0}(g) - (1/2\sqrt{6})[2 + \cos^2(\theta)(1 + 3\cos(2\psi))]A_{2,0}(g)$$

$$+ i\cos^2(\theta)\sin(2\psi)A_{2,1}(g) + (1/2)[2 + \cos^2(\theta)(1 - \cos(2\psi))]A_{2,2}(g). \quad (5.9)$$

Contributions in (5.9) not in (4.2) are permitted by the absence of selection rules from sites (2c) used in $P2_1'11$. The anapole does not feature in $F_{\pi'\sigma}$, even though it is an allowed multipole, and magnetic charge contributes through $A_{0,0}(g)$.

## 6. Conclusions

Magnetic properties of $LiMPO_4$ with M = Mn, Fe, Co and Ni have been discussed, with emphasis on properties visible in neutron and x-ray scattering. Particular attention is given to magnetic charges, or magnetic monopoles, that we predict in calculations of resonant x-ray Bragg diffraction amplitudes for $LiCoPO_4$ (monoclinic) and $LiNiPO_4$ (orthorhombic). Recent, independent *ab initio* simulations of the electronic structure of $LiNiPO_4$ support the existence of magnetic monopoles in this compound [14]. All being well, resonant x-ray Bragg diffraction experiments will soon be made on both the monoclinic and the orthorhombic compounds.

**Acknowledgement**

One of us is grateful to Dr K S Knight and A Rodríguez-Fernández for valuable discussions about the physical properties of lithium compounds with the olivine structure. In 2012 Dr V Scagnoli contributed in an initial calculation for $LiNiPO_4$.

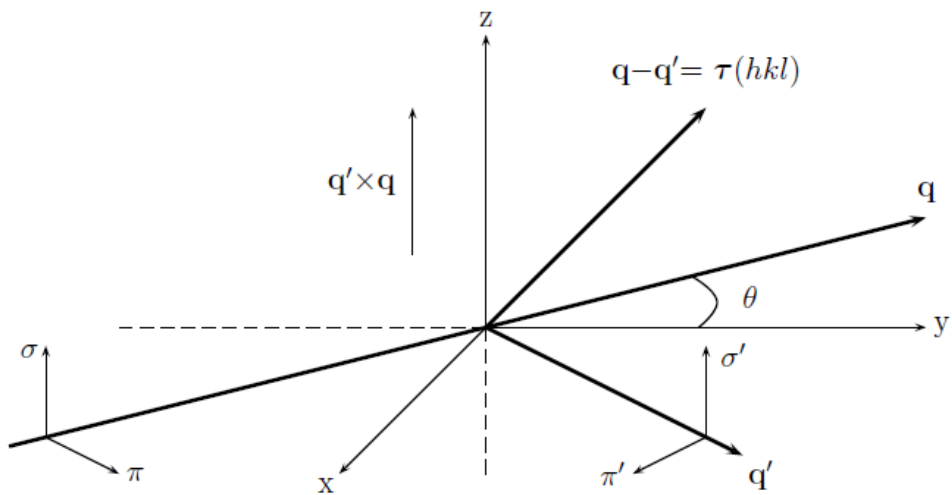

Figure 1. Cartesian coordinates (x, y, z) and x-ray polarization and wavevectors. The plane of scattering spanned by primary (**q**) and secondary (**q'**) wavevectors coincides with the x-y plane. Polarization labelled σ and σ' is normal to the plane and parallel to the z-axis, and polarization labelled π and π' lies in the plane of scattering.

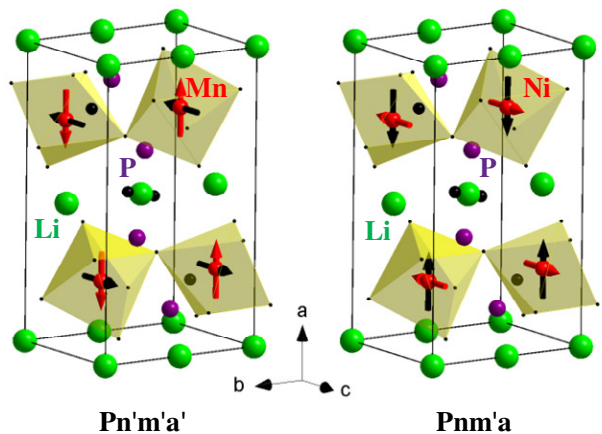

Fig. 2 Orthorhombic crystal structure of LiMPO$_4$ (M = Mn and Ni) and magnetic dipole motifs associated with $\Gamma^-_1$ (left) and $\Gamma^-_4$ (right) irreducible representations of the Pnma1' space group.

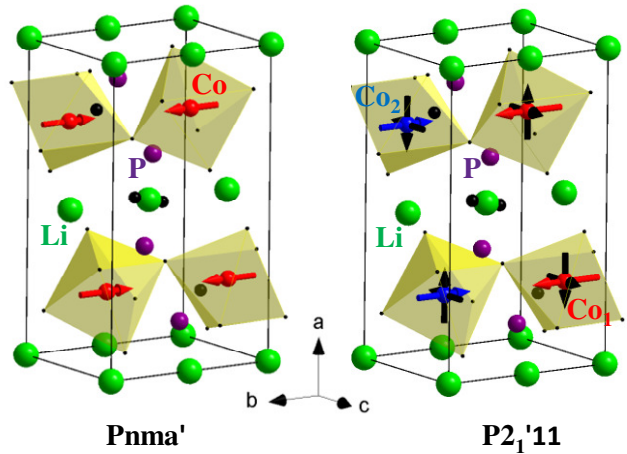

Fig. 3 Orthorhombic crystal structure of LiCoPO$_4$ and magnetic dipole motif associated with $\Gamma_2^-$ irreducible representation of the Pnma1' grey-group (left). Monoclinic crystal structure of LiMPO$_4$ and magnetic dipole motif associated with a $\Gamma_2$ irreducible representation of the P2$_1$11' grey-group (right). Two non-equivalent cobalt sites are denoted as Co$_1$ and Co$_2$, and symmetry places no constraint on relative orientations of magnetic dipoles in the two pairs. However, magnetic dipole moments associated with Co$_1$ and Co$_2$ nearly compensate, as shown in the cartoon, due to an antiferro-magnetic exchange between the two sub-lattices.